\documentclass[%
preprint, 11pt, titlepage, tightenlines,
superscriptaddress, 
nofootinbib,
nobibnotes,
amsmath,amssymb, 
showkeys,
]
{revtex4-2}
\usepackage{siunitx}
\usepackage{graphicx}
\usepackage{dcolumn}
\usepackage{bm}
\usepackage{bbm}
\usepackage{amsmath}
\usepackage[hidelinks]{hyperref}
\usepackage{gensymb}
\usepackage{xcolor}
\newcommand{\MoS}{MoS\textsubscript{2} }

\newcommand{\VBG}{$V_{\rm BG}$ }
\newcommand{\VDS}{$V_{\rm DS}$ }
\newcommand{\IDS}{$I_{\rm DS}$ }
\newcommand{\Supp}{Supporting Information }
\newcommand{\IV}{\textit{I}-\textit{V} }
\begin{document}
	\title{Addressing the spin-valley flavors in moir\'e mini-bands of MoS\textsubscript{2}} 
	\date{\today}
	\author{Chithra H. Sharma }%
	\email{Chithra.Sharma@physik.uni-hamburg.de}
	\affiliation{%
		Center for Hybrid Nanostructures (CHyN),  Universit\"at Hamburg,   Luruper Chaussee 149, Hamburg 22761 Germany. 
	}
	\affiliation{%
		Institut f\"ur Experimentelle und Angewandte Physik, Christian-Albrechts-Universit\"at zu Kiel, 24098 Kiel, Germany.
	}
	
	\author{Marta Prada}%
	\email{mprada@physnet.uni-hamburg.de}
	\affiliation{%
		I. Institute for Theoretical Physics,  Universit\"at Hamburg,   Luruper Chaussee 149, Hamburg 22761 Germany. 
	}
	
	\author{Jan-Hendrik Schmidt}
	\affiliation{%
		Center for Hybrid Nanostructures (CHyN),  Universit\"at Hamburg,   Luruper Chaussee 149, Hamburg 22761 Germany. 
	}
	\author{Isabel González Díaz-Palacio}
	\affiliation{%
		Center for Hybrid Nanostructures (CHyN),  Universit\"at Hamburg,   Luruper Chaussee 149, Hamburg 22761 Germany. 
	}
	\author{Tobias Stauber}
	\affiliation{%
		ICMM-CSIC, Sor Juana In\'es de la Cruz 3,  Madrid 28049 Spain.
	}
	\author{ Takashi Taniguchi }
	\affiliation{%
		International Center for Materials Nanoarchitectonics, National Institute for Materials Science,  1-1 Namiki, Tsukuba 305-0044, Japan.
	}
	\author{ Kenji Watanabe }
	\affiliation{%
		Research Center for Functional Materials, National Institute for Materials Science, 1-1 Namiki, Tsukuba 305-0044, Japan.
	}
	\author{Lars Tiemann}
	\affiliation{%
		Center for Hybrid Nanostructures (CHyN),  Universit\"at Hamburg,   Luruper Chaussee 149, Hamburg 22761 Germany. 
	}
	\author{Robert H. Blick}
	\affiliation{%
		Center for Hybrid Nanostructures (CHyN),  Universit\"at Hamburg,   Luruper Chaussee 149, Hamburg 22761 Germany. 
	}
	\affiliation{%
		Material Science and Engineering, University of Wisconsin-Madison, University Ave. 1550, Madison, 53706, Wisconsin, USA.
	}
	%
	\begin{abstract} 
		The physics of moir\'e superlattices and the resulting formation of mini-bands in van der Waals materials have opened up an exciting new field in condensed matter physics. These systems exhibit a rich phase diagram of novel physical phenomena and exotic correlated phases that emerge in the low-dispersing bands. Transition metal dichalcogenides, in particular, molybdenum disulfide (MoS$\textsubscript{2}$), are potential candidates to extend the studies on moir\'e electronics beyond graphene. Our transport spectroscopy measurements and analysis reveal a correlation-driven phase transition and the emergence of discrete mini-bands in MoS$\textsubscript{2}$ moiré superlattices that remained elusive so far. 
		We resolve these mini-bands arising from quantum mechanical tunneling through Schottky barriers between the MoS$\textsubscript{2}$ and its metallic leads. Energy scales deduced from a first approach exhibit an astounding agreement with our experimental observations. The behavior under thermal activation suggests a Lifshitz phase transition at low temperatures that is driven by a complete spin-valley symmetry breaking. These intriguing observations bring out the potential of twisted MoS$\textsubscript{2}$ to explore correlated electron states and associated physics.
	\end{abstract}
	%
	%
	%
	\maketitle
	%
	The periodic potential of a crystal structure induces bands of allowed energies that ultimately determine the physical properties of a solid. Moir\'e patterns, usually an ornate artifact of textured images, can also be created in atomically thin layers of van der Waals (vdW) materials inducing additional periodicity. The skillful placement of two vdW layers under a specific rotation angle leads to a moir\'e superlattice (MSL), resulting in unexpected and amazing physical properties. The first observations of these 'magic' changes to the physical properties were made in twisted bilayer graphene that suddenly exhibited a sequence of correlated insulating \cite{Cao2018}, superconducting \cite{Cao2018a, Yankowitz2019}, and ferromagnetic phases \cite{sharpe2019,serlin2020}. It is a general consensus that a reconstruction of the energy dispersion flattens the electronic bands, essentially turning massless Dirac fermions into heavily interacting particles \cite{Bistritzer11}. These discoveries in graphene triggered a run to the extended family of vdW transition metal dichalcogenides (TMDCs) \cite{Cai2021, Wang20, Wu2018, Pan2020, Devakul2021,Naik2018, Naik2020, Xian2021}. 
	
	The primary advantage of TMDCs over graphene is that the stringent magic angle condition for flat band formation is relaxed \cite{Wu2018, Pan2020}. It did not take long until experimental reports of Mott insulating phases \cite{Wang20} followed for magic angle TMDCs. The attention has recently turned to flat-band related phenomena at zero magnetic field, where spontaneous symmetry breaking \cite{Lemonik10} causes a cascade of Lifshitz transitions \cite{Zondiner20,Barrera22, Wong20}. These transitions occur in bilayer graphene as the spin and valley degeneracy are spontaneously broken at integer values of moir\'e band filling.  Owing to its abundance and stability, an attractive TMDC candidate to observe these phenomena is the conduction band of twisted molybdenum disulfide (MoS\textsubscript{2}) structures \cite{Mak2010}. Although the moir\'e bands were theoretically predicted \cite{Naik2018, Naik2020} and the small spin-orbit coupling (SOC) \cite{Marinov17} ensures quasi-degenerate states, the MSL in MoS\textsubscript{2} is relatively unexplored in experiments and has not been attempted so far. 
	
\begin{figure*}[!hbt]
		\centering
		\includegraphics{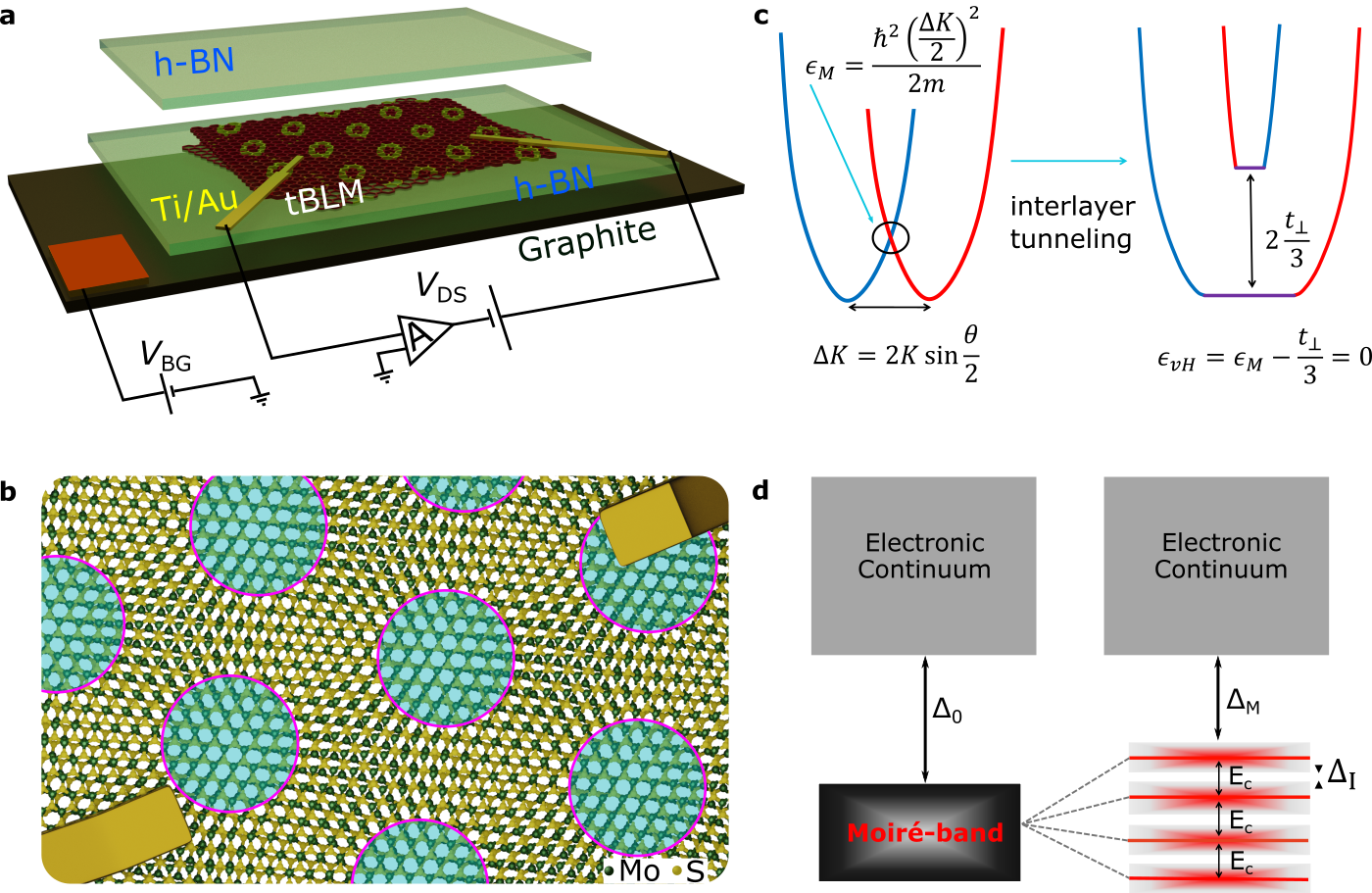}
		\caption[]{
			\textbf{a}, Schematics of the sample configuration with the moir\'e superlattices encapsulated in hBN. \textbf{b}, Schematic showing the atomic arrangement of a twisted bilayer MoS\textsubscript{2}, with AB stacked domains marked in pink circles. 
			\textbf{c}, Energy dispersion and \textbf{d}, band diagrams illustrating the band hybridization near the magic angle or flat band condition (left) and the spontaneous spin-valley symmetry breaking that emerges with interactions (right).
		}
		\label{fig_theory}
	\end{figure*}

	In this work, we experimentally resolve and explore the discrete mini-bands in MSLs of twisted MoS\textsubscript{2} structures encapsulated in hexagonal boron nitride (hBN) using current-voltage (\textit{I}-\textit{V}) characteristic measurements. A schematic of the device and experimental circuiting is depicted in Fig. \ref{fig_theory}a. A graphite electrode at the bottom works as a back-gate and hBN as a dielectric. The electrical contacts are represented by the two golden stripes. Fig. \ref{fig_theory}b shows a cartoon of the MSL in MoS\textsubscript{2}. We exploit the negative differential conductance (NDC), induced by the Schottky barriers (SB) at the electrical contacts, to resolve four mini-bands arising from the spin-valley symmetry breaking and corresponding energy gaps in the proximity of the conduction band of twisted \MoS structures. 
	
	\subsection*{Theoretical background:}
	Our experimental data discussed below agree with a simple model that considers the lifting of spin and valley degeneracy in the magic-angle moir\'e band, as observed in graphene-based structures \cite{Zondiner20, Wong20, Zhou22, Barrera22,Cao20, Liu20b, Zhou21} and twisted bilayer TMDCs \cite{Wang20}. The normally four-fold degenerate conduction band of the individual layers that form the  moir\'e system has a parabolic dispersion $\epsilon({\bf k})=\frac{\hbar^2k^2}{2m}$, where $m=0.76 m_0$ with $m_0$ the free electron mass \cite{Rostami13}. Due to the twist angle, the minima of the parabolic dispersion of the two layers are separated by the wavevector $\Delta K=2K\sin(\theta/2)$, where $|{\bf K}|=K=\frac{4\pi}{3a}$ as shown in the left of Fig. \ref{fig_theory}a \cite{Rostami13}. The parabolic bands of the upper and lower layer anti-cross at the so-called moir\'e energy $\epsilon_M=\epsilon(\Delta K/2)$. Band hybridization via interlayer hopping $t_\perp\simeq(40-50)$ meV \cite{Paradisanos2020,Gong13} results in spin and valley degenerated bonding and anti-bonding states, as shown schematically on the right-hand side of Fig. \ref{fig_theory}c. The magic angle condition reads \cite{Stauber_2013}:  
	
	\begin{align}
		\epsilon_M-\frac{2t_\perp}{3}=0\;. 
	\end{align} 
	Here, the gap enhancement owing to lattice relaxation is accounted for by adding a factor 2 above \cite{Nam17,Koshino18}. This gives a magic angle of 
	$\theta_M\simeq\frac{3a}{2\pi}\sqrt{\frac{2m}{\hbar^2}\frac{t_\perp}{3}}\approx (4-5)^\circ$. Near the magic angle, the single-particle spectrum of the low SOC conduction band can thus be approximated by a valley and spin four-fold degenerate ground-state that forms a mini-gap of $\Delta_M \simeq 2t_\perp/3$ with the continuum.
	
	In the presence of interactions, this four-fold degeneracy is broken \cite{Stauber22}. Simple order of magnitude arguments show that the leading electron-electron interaction in the system is the long-range Coulomb interaction. This gives an estimate of the Coulomb energy to be of the order of $E_{\rm c} = e^2/(4\pi\epsilon\epsilon_0 L) \simeq (40-50) $ meV, where $\epsilon_0$ is the permittivity of free space, the dielectric constant, $\epsilon \simeq (4-6)$ is the screening from the hBN environment \cite{Laturia2018} in our samples and $L \sim a/2\sin(\theta_m/2) \sim 5$ nm  \cite{Lopes07}, with $a$ = $0.319$ nm being the lattice constant. At integer moir\'e band fillings, the electron-electron interaction results in incompressible states \cite{Zondiner20} causing Lifshitz phase transitions and a band structure as schematically depicted in Fig. \ref{fig_theory}d: Whenever the occupation per moir\'e cell reaches an integer value, the corresponding virtually occupied channel splits up by the Coulomb energy, $E_{\rm c}$.
	
	
	\subsection*{Experimental data and analysis} 
\begin{figure*}[!hbt]
		\centering
		\includegraphics{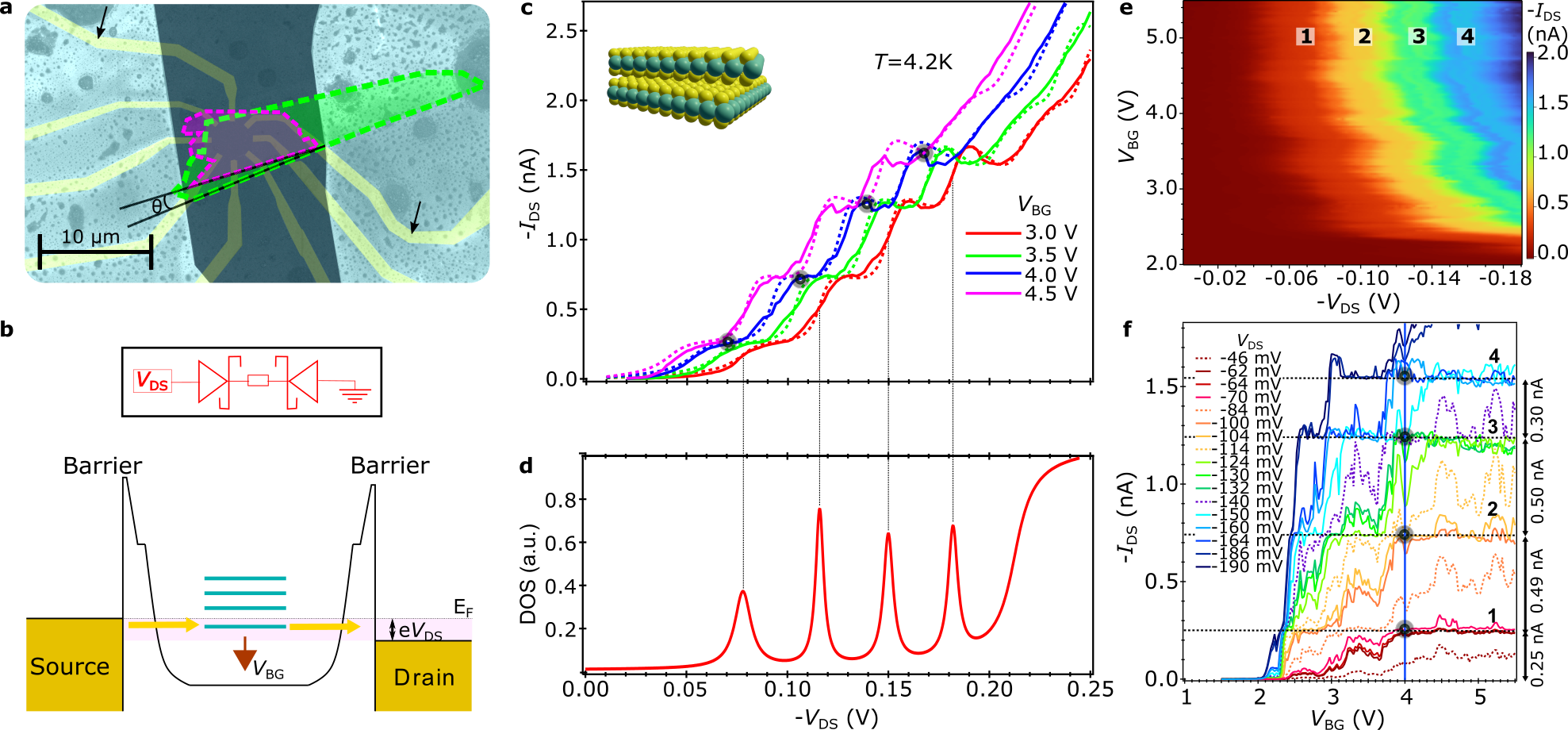}
		\caption{Resonant tunneling and negative differential conductance in tBLM (device A). 
			\textbf{a}, False-colored SEM image of the device with the two \MoS flakes outlined in green and pink for clarity. The two contacts (source and drain) used for the measurement are marked with black arrows (see \Supp section S1 for more details). \textbf{b}, a Level diagram illustrating transport via discrete moir\'e mini-bands. The inset shows an equivalent circuit diagram with the Schottky contact represented as a diode.
			\textbf{c}, Experimental \IV traces (solid lines) and theoretical fits (dashed) for different back-gate voltages, $V_{\rm BG}$. The inset is a cartoon representing the stacked bilayer with a twist. 
			\textbf{d}, DOS extracted (see Supplementary Information, section S4) from the \IV traces, plotted for \VBG = 3 V. 
			\textbf{e}, Contour plot of the current as a function of \VDS and \VBG depicting the four plateaus. 
			\textbf{f}, Current as a function of back-gate voltage, showing step-like, non-linear behavior for different $V_{\rm DS}$.  A blue vertical line at $V_{\rm BG}$ = 4 V corresponds to the blue curve in \textbf{a} and the black dots on both plots correspond to the plateau regions with the same parameters. Horizontal lines are guides to the eye for values where the current flattens. 
		}
		\label{figIVs}
	\end{figure*}

	We investigate a twisted  bilayer MoS\textsubscript{2} (tBLM) (device A) and  two twisted double-bilayer MoS\textsubscript{2} (tDBLM) (devices B and C). The tBLM (tDBLM) were fabricated by aligning two monolayers (bilayer) MoS\textsubscript{2} flakes along their straight edges and then rotating the upper one by a small angle $\theta$. A home-built vdW aligner setup, equipped with translation and rotation stages along with a microscope, was used for the stacking. Fig. \ref{figIVs}a shows an SEM image of device A, false-coloured for clarity. The perimeter of the lower and upper monolayers MoS\textsubscript{2} flakes are marked with green and pink dashed lines, respectively. The long straight edges of the flakes are assumed to be zig-zag, which is the preferable cleavage direction for MoS\textsubscript{2} \cite{Guo2016, Wang2015}. The angle $\theta$ between the straight edges of the monolayers (bilayers) regions is deduced from optical images, as illustrated in  Fig. \ref{figIVs}a. Further details on sample fabrication and angle determination are provided in methods and \Supp S1. Owing to the point group symmetry of MoS\textsubscript{2}, we may assume an angle of $\theta$ or $60\degree\pm\theta$ (see \Supp S2 for details). Table \ref{table1} summarizes the measured relative twist-angles for the three devices discussed further in this manuscript. A small error in this measurement due to the optical resolution limit is not considered here. 
	\begingroup
	\begin{table}[!h]
		\caption{Relative twist angle of devices A, B and C.}
		\label{table1}
		\begin{ruledtabular}
			\begin{tabular}{l|c|r} 
				\textbf{Device} & \textbf{Type} & \textbf{$\theta$}\\
				\hline
				A & tBLM & $4.40\degree$\\
				B & tDBLM & $6.22\degree$\\
				C & tDBLM & $1.21\degree$\\
			\end{tabular}
		\end{ruledtabular}
	\end{table}
	\endgroup

	Ti/Au (5 nm/15 nm) contacts are defined on the twisted stacks by electron beam lithography, metallization and lift-off. The contacts are marked by golden stripes in Fig. \ref{figIVs}a. Ti/Au contacts on MoS\textsubscript{2} are known to form Schottky contacts which are usually detrimental for electrical transport  \cite{Kaushik2014, Gong2014}. Here, however, we exploit the quantum mechanical tunneling through the barrier to resolve the discrete MSL mini-bands as detailed below.  We model the Schottky contacts as diodes symmetrically connected as in the circuit representation shown in the top-inset of Fig. \ref{figIVs}b. The transport at the barriers allows us to address the mini-bands as illustrated in Fig. \ref{figIVs}b.  All measurements were performed by applying a DC drain-source bias voltage ($V_{\rm DS}$) in a two-probe configuration. The current ($I_{\rm DS}$) is then measured through a current pre-amplifier, which also acts as a virtual ground.  We apply a DC voltage at the graphite back-gate ($V_{\rm BG}$) to shift the Fermi level through the discrete mini-bands. Device A  was measured at 4.2 K and devices B and C at 1.5 K, unless stated otherwise. 
	
	The \IV characteristics for device A in reverse (negative) bias operation are shown in Fig. \ref{figIVs}c for selected gate voltages above the on-state,  \VBG$>3$ V. The solid lines are experimental data, whereas the dashed lines correspond to theoretical fits (see methods). For low bias $|V_{\rm DS}| < 20$ mV a negligible current flows through the device, due to the absence of conducting channels at the Fermi-level ($E_F$). Increasing $|V_{\rm DS}|$ boosts the tunneling current \cite{Yu1998}, as bands are shifted into the conducting window near the Fermi level. The four step-like structures in Fig.  \ref{figIVs}c with increasing bias reveal the discrete nature of the moir\'e bands below the conduction band \cite{Naik2018, Naik2020, Maity2020}. As the gate voltage increases, the MSL levels shift down in energy (see diagram in Fig. \ref{figIVs}b),  lowering the bias threshold $|V_{\rm DS}|$. The steps shift linearly with $V_{\rm BG}$, indicating that, for small biases used here, the gate affects the levels homogeneously and independently of the barrier. Note that such a linear shift is typically observed in resonant tunneling transistors \cite{linh2014,Reed88}. 
	
	A salient feature in Fig. \ref{figIVs}c is the observation of a negative differential conductance (NDC) region following each step in the  \IV curve. A possible explanation would be that the bias increases the SB height \cite{Esaki1970} while no new states are added for conduction, resulting in NDC. However, correlation effects can also point to NDC, owing to the formation of a bound state \cite{Ishigaki2020,Shin1998}, which is sensitive to temperature \cite{Nguyen2004}. We stress that these features are absent in non-twisted samples. \Supp S3 shows \IV measurements from a few-layer MoS\textsubscript{2} device for various \VBG at 4.2 K. 
	
	Employing a simple Landauer formalism and assuming a Breit-Wigner density-of-states (DOS), we were able to fit the experimental data (broken lines of Fig. \ref{figIVs}c) \cite{Breit1936}. A plot of the extracted DOS as a function of \VDS is shown in Fig. \ref{figIVs}d. (See methods below and \Supp S4 for details). The theoretical Coulomb energy $E_{\rm c} \simeq$ (40-50) meV that we deduced earlier is strikingly close to the corresponding bias separation of the steps observed in Fig. \ref{figIVs}c. From the extracted DOS (see Fig. \ref{figIVs}d), we obtain a value for the charging energy of $E_{\rm c} \simeq$ (32 $ \pm$ 2) meV and a mini-gap of $\Delta_M \simeq$ (38 $\pm$ 5)  meV, which are illustrated in Fig. \ref{fig_theory}d. 
	
	\begin{figure*}[!hbt]
		\centering
		\includegraphics{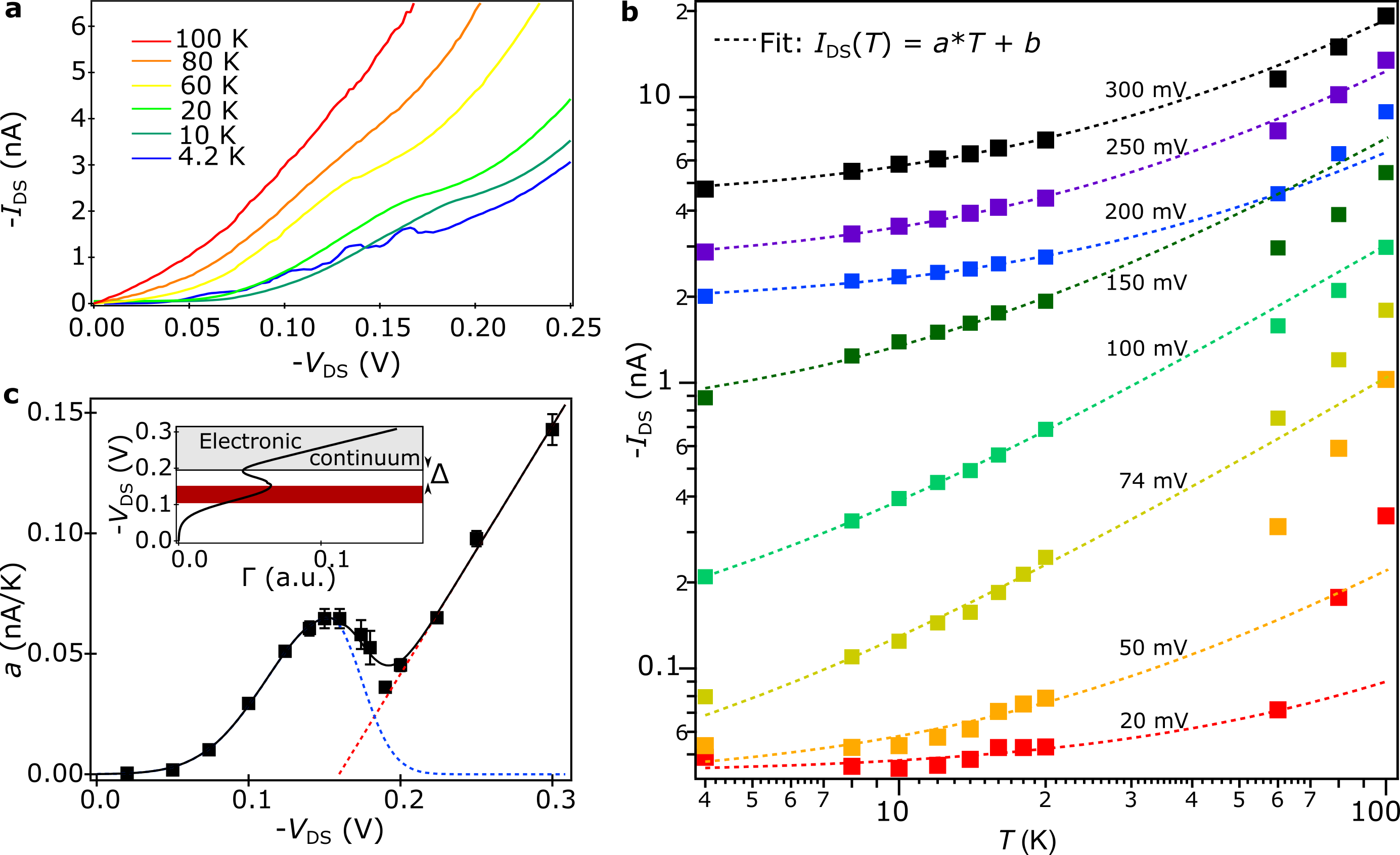}
		\caption{ \textbf{a}, Experimental \IV traces of device A for different temperatures at $V_{\rm BG}$ = 4 V and 
			\textbf{b}, Log-log plot of the current as a function of temperature for different biases. The linear fit (broken lines) to the experimental data (squares) is well suited at low temperatures, $T\lesssim 20$ K. 
			\textbf{c}, The linear coefficient $a$ (related to the transmission coefficient, $\Gamma$) which is sensitive to the sample's DOS plotted against $V_{\rm DS}$: The peak at $\sim$ 150 mV relates to the moir\'e band, whereas the linear increase corresponds to the constant DOS or the electronic continuum in the 2D system, as illustrated in the inset. 
		}
		\label{figIVTemp}
	\end{figure*}
	
	Fig. \ref{figIVs}e shows a contour plot of  $I_{\rm DS}$ as a function of $V_{\rm DS}$ and $V_{\rm BG}$. The emergence of clear plateaus in the current allows us to resolve discrete levels marked by 1 through 4. These four clear plateaus in the current are evidence of a non-trivial energetic spectrum.  We stress, however,  that the SB prevent us from extracting the conductance in quantized units of $e^2/h$. Fig. \ref{figIVs}f shows line cuts with the current as a function of \VBG  at different $V_{\rm DS}$.  The blue vertical line at \VBG= 4 V corresponds to the blue curve in Fig. \ref{figIVs}c and the black dots on both plots correspond to the plateau regions with identical parameters. The current flattens to four well-defined values of separation (0.25 $\pm$ 0.02) nA, (0.49 $\pm$ 0.02) nA, (0.50 $\pm$ 0.02) nA and (0.30 $\pm$ 0.02) nA, for particular windows of drain-source and gate voltages, marked by the horizontal lines (guides to the eye). Remarkably, the current exhibits a larger amount of noise for drain-source bias between these windows (in dashed lines), and well-defined, low noise values for bias within the windows. See \Supp S5 for the noise spectra extracted from the data. The second and third steps show roughly an increment of twice the step size as the first and fourth. We speculate  that this could be related to the nature of the different broken symmetries (spin and valley), or the non-trivial connection of the bands \cite{Lemonik10}.  A deeper theoretical analysis beyond this experimental work would be needed to unravel further details of this low-noise region and the different shapes of the even and odd steps. 
	
	It is worth noting that the lifting of spin-valley symmetry requires energies of the order of $k_B T \lesssim$ (1-2) meV, corresponding to the energy scale at which the moir\'e subbands split in presence of long -range interaction \cite{Stauber22, Balents2020}. Fig. \ref{figIVTemp}a. shows the temperature dependence of the \textit{I-V} characteristics at \VBG= 4 V. The transition is indeed lost as the temperature increases by only a few K, serving as an indication of the energy scale to observe the  correlated state \cite{Balents2020}. We note that the sample was re-cooled for the temperature-dependent measurements. Consequently, an increase in the SB causes the crossing of the blue and green curves at low bias voltages. The remaining step is washed out at $\sim$ 100 K ($k_BT \simeq$ 9 meV), corresponding to the energy scale of the SB height, where the linear approximation breaks down.
	
	\begin{figure*}[!hbt]
		\centering
		\includegraphics{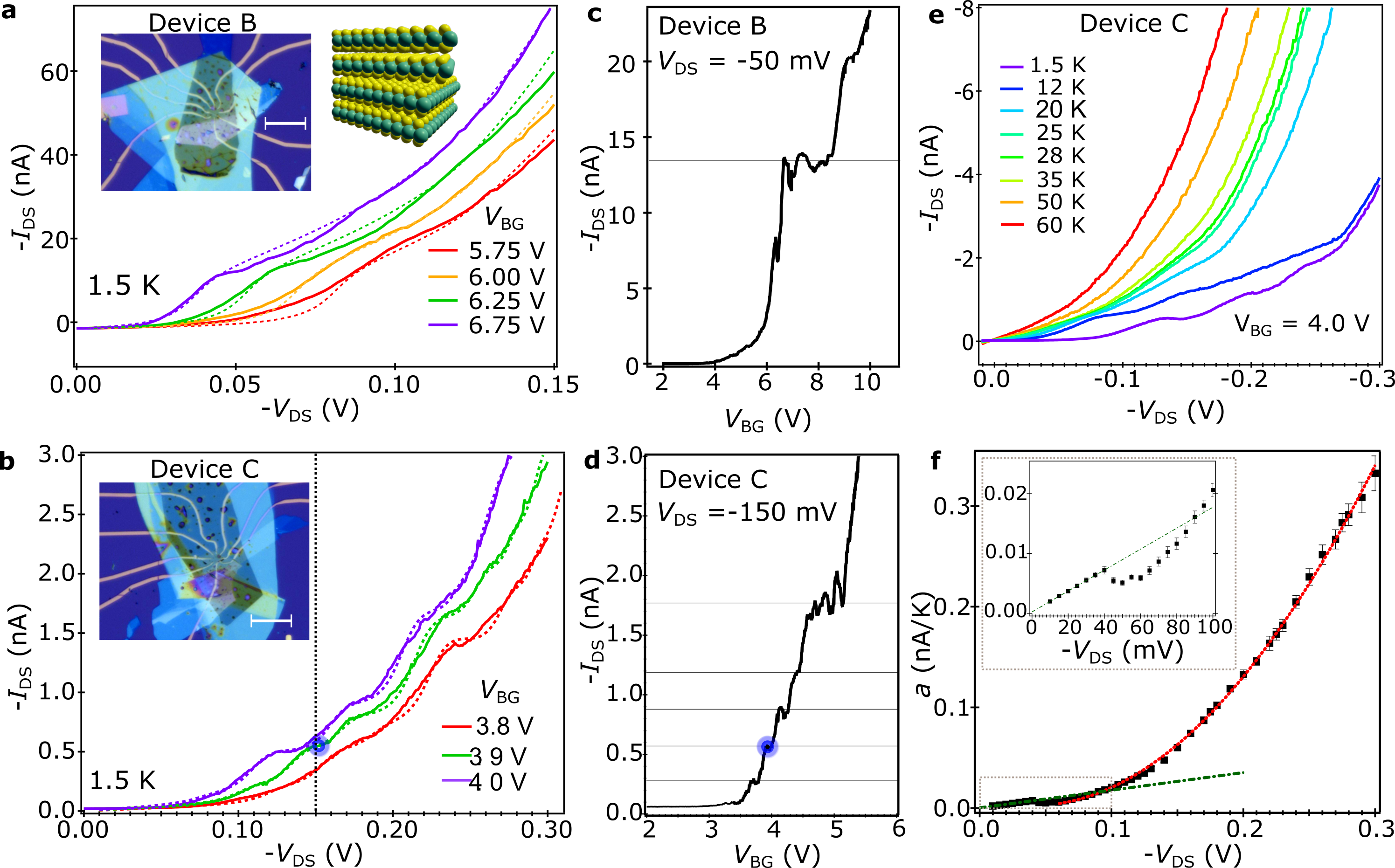}
		\caption[]{
			\textbf{a} and \textbf{b}, \IV characteristics for different \VBG for devices B and C, respectively. The optical images of the respective devices are shown in the inset. The scale bar is 20$\mu$m. The right inset in panel \textbf{a} depicts the double-bilayer structure. The angle between the upper and lower bilayers is 6.22 $\pm$ 0.88\degree (device B) and 1.21 $\pm$ 1.74\degree (device C). 
			\textbf{b} and \textbf{d}, Current as a function of back-gate voltage, showing step-like, non-linear behavior for devices B and C, respectively. The black horizontal lines in \textbf{d} correspond to the plateau regions and the blue dots in \textbf{b} and \textbf{d} correspond to the same parameters. 
			\textbf{e}, Temperature dependence of the \IV characteristics for device C. \textbf{f}, linear coefficient extracted from the temperature dependence as in \ref{figIVTemp}c.
		}
		\label{figTDBLM}
	\end{figure*}
	
	The current as a function of temperature for different biases is depicted in Fig. \ref{figIVTemp}b employing a log-log scale. At low temperatures, $T\lesssim 20$ K, the current increases linearly with temperature, $I_{DS} = a(V_{ \rm DS}) T + b$, a common feature of tunneling through SB \cite{Yu1998}. The dashed curves correspond to the linear fit, whereas the filled squares are experimental data points. The extracted linear coefficient is then represented as a function of \VDS in Fig. \ref{figIVTemp}c. Two regimes can be distinguished: A Lorentzian around 160 mV (blue broken curve), corresponding to the  Breit-Wigner shaped DOS of the moir\'e band, and a linear increase (red line), which can be identified with the constant DOS of the electronic continuum. Note that $ a(V_{DS})$ is closely related to the quantum transmission coefficient of electrons below the barrier (see \Supp S4 for details) as shown in the inset. The black curve corresponds to the total transmission, which is the sum of the red and blue traces. 
	
	Next, we consider two double-bilayer structures. The \IV characteristics in reverse (negative) bias operation for devices B and C at positive back-gate voltages are shown in Fig. \ref{figTDBLM}a and \ref{figTDBLM}b, respectively. The  main difference  between these two samples is the relative angle between the top and bottom bilayer, which is 6.22$\degree$ for device B and 1.21$\degree$ for device C. The solid lines are experimental data, whereas the dashed lines correspond to a theoretical fit (see methods). While only one  mini-band is resolved for sample B, sample C shows a set of mini-bands. This is consistent with the picture where a larger MSL (smaller angle) causes the favorable stacking region to occupy the largest area, lowering the symmetry of the MSL and favoring the appearance of moir\'e bands with larger confinement \cite{Naik2020}. Note that, in contrast to device A, devices B and C have larger \IDS and the shift in the steps in \IV as a function of \VBG shows a non-linear trend. This could be related to the non-uniformity of the vertical current flowing through a multiple-layered structure  \cite{das2013}, preventing us to resolve the mini-gaps for some bias. However, we determine a Coulomb energy of $E_{\bf c} \simeq$ (32 $\pm$ 2)  meV, which is consistent with the larger MSL structure. 
	
	In order to discriminate the nature of the features of Fig. \ref{figTDBLM}b, we evaluate the \IV characteristics as a function of the temperature. As we can see in Fig. \ref{figTDBLM}e, the features disappear around 25 K.  We observe a linear behavior of the transmission with the temperature up to  60 K, revealing a higher energy scaling of the SB. The coefficient shows also a non-monotonic behavior at around 50 meV, which we identify with the moir\'e band. This behavior is consistent with the observation of temperature dependence in device A.  This is a strong indication that the origin of the mini-bands is indeed due to the breaking of the spin-valley degeneracy.
	
	In conclusion, our observation of mini-bands in MSLs of MoS\textsubscript{2} opens up a new platform to study correlated electrons in TMDCs and a new direction in the field of twistronics. Though mini-bands are observed in other TMDCs, their observation in MoS\textsubscript{2} in transport has not been reported so far despite various theoretical predictions. While the formation of Schottky contacts in MoS\textsubscript{2} is considered unfavorable to observing correlated electron states, we harness the same to observe discrete energy states in the conduction band from the measurement of \IV characteristics. From theoretical fits to the \textit{I}-\textit{V}, we can resolve the energy levels separated by $\sim$32 meV for twisted bilayer devices. In spite of the large gaps, the steps in the \textit{I}-\textit{V}s could only be observed at very low temperatures. This suggests that their origin is related to electron correlation where complete lifting of the spin-valley states induces a phase transition. This is facilitated by the low SOC in the conduction band of \MoS, implying a quasi 4-fold degeneracy of the charge carriers. Similar phase transitions were also seen in magic-angle twisted bi-layer graphene \cite{Zondiner20,Barrera22, Wong20}.  We also observe similar features in twisted double-bilayer structures consistent with our inferences. We believe that our results motivate further experiments on MSLs of MoS\textsubscript{2}, especially achieving Ohmic contacts and precision in twist-angle will be promising.
	
	
	\section*{Methods}\label{SecMeth}
	\textbf{Device fabrication.}
	Commercially procured natural MoS\textsubscript{2} (SPI supplies) were used for our studies. We fabricate the vdW stack by dry-transfer method with Poly-dimethyl-siloxane (PDMS) using a homemade aligner setup. The setup has a fine rotation stage and heating to facilitate the fabrication of the twisted bilayer devices. During the transfer, the stage is heated to 70$\degree$C for five minutes to reduce wrinkles and remove trapped air bubbles during the process. The flakes are exfoliated from the bulk on a PDMS sheet and micro-positioned and transferred to a Si/SiO$_2$ wafer. The work-flow and twist angle determination are described in detail in the \Supp S1. \\
	The thickness of the hBN at the back-gate for device A is estimated to be (27 $\pm$ 3) nm from AFM measurements. 
	
	\textbf{Transport measurements.}
	The devices were mounted on a chip carrier in a vacuum (2$\times$10$^{-5}$ mbar) probe. Device A was cooled to 4.2 K in a bath cryostat and devices B and C were cooled to 1.5 K in a VTI. All measurements were performed in dark. Temperature-dependent measurements on device A were performed in a Quantum Design PPMS. The device was warmed and re-cooled for this measurement. Temperature dependence of device C was done in the VTI by regulating the helium flow and pressure in the insert. For all the measurements, a DC voltage was applied across the source-drain contacts and the \IDS was measured using a current pre-amplifier. A source-measure unit was used for applying DC voltage to \VBG and the leakage current was monitored to ensure that no current is leaking to the gates. 
	
	\textbf{Numerical fit.}
	We employ a Landauer formalism to account for the tunneling current through the contacts and sample operating in reverse bias \cite{daviesbook}, 
	\begin{equation}
		\label{eqI}
		I \propto\int_0^{\infty} 
		\Gamma (E) {\rm DOS} (E) e^{-(E-E_{Fs})/k_{\rm B}T}dE
	\end{equation}
	where $ {\rm DOS}(E)$ represents the density of states in the two-dimensional electron gas, for which a Breit-Wigner distribution is employed for 
	discrete bands and $T(E)$ is the transmission over the tunneling barrier. We assume a triangular barrier for the Schottky contact \cite{Yu1998} within the WKB approach, $T(E) \propto \exp[{-( e\phi_i+eV-E)/E_0}]$. The effects of the gate are then included by shifting the energy scale on the DOS by $e\alpha V_{BG}$, where $\alpha$ is a rational function of the capacitance of the system \cite{Beenakker1991}. We extract $\alpha = 0.025$, which is reasonable for a gate distance of 27 nm. The theoretical fit (dashed lines of Figs. 2 and 4) are obtained by numerical integration of (\ref{eqI}), where the only free parameter is the barrier height, $\phi_i$. More details are included in the \Supp S4.  
	
	\section*{Data Availability}
	Data that support the findings of this study are available from the corresponding authors upon reasonable request.

	\section*{Acknowledgments}
	The authors thank Dr. Robert Zierold for the usage of PPMS, Dr. Pai Zhao, Jun Peng and Vincent Strenzke  for their help with the experiments and critical reading, Prof. Kai Rossnagel, Prof. Madhu Thalakulam, Prof. Tim Wehling, Prof. Gabriel Bester and Carl Nielsen for discussions and suggestions.
	
	The work was supported by the DFG (project MEGA-JJ, Grant No. BL-487/14-1)  and PIER seed fund (CorMoS, Grant No. PIF-2021-01).
	C.H.S acknowledges the Alexander von Humboldt foundation for post-doctoral research fellowship.
	I.G.P acknowledges the BMBF "SMART" project (Grant No. 05K19GU). 
	T.S. was supported by the project No. PID2020-113164GB-I00 financed by MCIN/AEI/10.13039/501100011033.
	K.W. and T.T. acknowledge support from the JSPS KAKENHI (Grant Nos. 19H05790, 20H00354 and 21H05233).
	
	\section*{Competing interests}
	The authors declare no competing interests
	
	\section*{Additional Information}
	\Supp

\end{document}